\documentclass[twocolumn,showpacs,preprintnumbers,amsmath,amssymb]{revtex4}
\usepackage{graphicx}
\usepackage{dcolumn}
\usepackage{bm}

\begin{document}

\title{Electric Field Effects and the\\  
Experimental Value of the Muon g-2 Anomaly}

\author{A. Widom}
\author{Y.N. Srivastava}
\affiliation{Physics Department, Northeastern University, Boston MA 02115}
\affiliation{Physics Department \& INFN, University of Perugia, Perugia Italy}

\begin{abstract}
The electric field corrections to the recently measured muon 
magnetic moment g-2 anomaly are considered from both the classical 
(BMT) and the quantum mechanical (Dirac) viewpoints. In both views,  
we prove that the electric field inducing the horizontal betatron 
tune does not renormalize the anomaly frequency. With this result 
kept in mind, the experimental muon magnetic moment anomaly 
is in closer agreement with standard model predictions than 
has been previously reported.  
\end{abstract}

\pacs{29.20.Fj, 29.27.Bd, 29.27.Fh, 13.35.Bv, 13.30.Em}
\maketitle

There has been considerable experimental effort in accurately 
determining\cite{1,2,3,4,5,6,7,8} the muon magnetic moment anomalous 
\begin{math} g \end{math}-factor 
\begin{equation}
\kappa =(g-2)/2 .
\end{equation}
The measurement is performed via the magnetic field induced anomaly 
frequency 
\begin{equation}
\omega_\kappa =\left|{e\kappa {\bf B}\over Mc}\right|.
\end{equation}

The increased precision of the most recent experimental 
determinations\cite{7,8} of \begin{math} \kappa  \end{math} 
is mainly due to the technical progress that 
has been made in producing a uniform magnetic field 
\begin{math} {\bf B} \end{math} which is thought to 
yield a very accurately determined frequency
\begin{math} \nu =(\omega_\kappa /2\pi ) \end{math}. It has been 
claimed \cite{7,8} that the anomaly \begin{math} \nu \end{math} 
is known to within 
\begin{math} \Delta \nu \sim 5\times 10^{-2}\ Hz \end{math} by signal 
averaging over experimental runs with finite (due to muon decay) 
time intervals obeying    
\begin{math} \Delta t\sim 7\times 10^{-3}\ sec \end{math} .  
Thus, there is an experimental uncertainty product  
\begin{math} \Delta \nu \Delta t \sim 4\times 10^{-4}<<1 \end{math}. 
With \begin{math} N\sim 3\times 10^9  \end{math} muon decay 
events, the reported frequency resolution also obeys 
\begin{math} 
(\delta \nu /\nu )\sim 10^{-6}<1/\sqrt{N}\sim 2\times 10^{-5} 
\end{math}.

For a muon with kinetic energy 
\begin{math} {\cal E}=Mc^2\gamma \end{math},  
it has been proposed from a classical spin viewpoint\cite{5,6,7,8}, 
that the anomaly frequency Eq.(2) has an additional electric field 
induced shift \begin{math} \delta \omega_{\perp \kappa }\end{math} given by  
\begin{equation}
{\delta \omega_{\perp \kappa} \over \omega_\kappa }=
\left<\left\{{1\over \kappa (\gamma^2-1)}-1\right\}
{({\bf v\times E})\cdot {\bf B}\over c|{\bf B}|^2}\right>,
\end{equation}
where \begin{math} <...> \end{math} denotes an average over the 
cyclotron orbits of the muon beam. To minimize the electric field 
effect, one attempts to choose the energy so that the term within 
brackets \begin{math} \{...\} \end{math} is as small in absolute 
magnitude as is experimentally possible.

The reported discrepancies\cite{5,6,7,8} between the  
standard model and experimental values of 
\begin{math} \kappa \end{math} are in part due to 
the overestimate of experimental electric field effects  
and the underestimate of standard model values and errors therein. 
When \begin{math} \delta \omega_\kappa \end{math} 
is properly computed, the experimental determination of 
\begin{math} \kappa  \end{math} must be evaluated anew. 
We find that the experimental value is thereby drawn closer to 
the standard model value.

As we show in what follows, the contribution to the anomaly 
frequency 
\begin{math} \delta \omega_{\perp \kappa } \end{math}
due to the electric field 
\begin{math} {\bf E}_\perp  \end{math} (perpendicular to 
\begin{math} {\bf B}  \end{math}) vanishes. The rule 
\begin{math} \delta \omega_{\perp \kappa}=0 \end{math} is most easily 
proved employing quantum mechanics via the Dirac equation. However 
the result may also be understood from a classical viewpoint. 
Both the quantum and classical arguments will be discussed in 
what follows. 

To evaluate Eq.(3) from a classical viewpoint, let us discuss the 
muon motion projected on to a plane perpendicular to the uniform 
magnetic field \begin{math} {\bf B}  \end{math}; 
i.e. let us define the position and velocity, respectively,  
\begin{equation}
{\bf r}_\perp 
=\left({{\bf B\times }({\bf r \times B})\over |{\bf B}|^2}\right)
\ \ {\rm and}\ \ {\bf v}_\perp 
=\left({{\bf B\times }({\bf v \times B})\over |{\bf B}|^2}\right).
\end{equation}
Employing the Lorentz force on the muon charge 
\begin{equation}
M{d(\gamma {\bf v})\over dt}=e\big({\bf E}+({\bf v\times B})/c\big),
\end{equation}
with 
\begin{equation}
{\bf \rho }={c\over e}
\left({{\bf B\times }(M\gamma {\bf v})\over |{\bf B}|^2}\right) 
\ \ {\rm and}\ \ {\bf r}_\perp ={\bf \rho }+{\bf R},
\end{equation}
we have in virtue of Eqs.(5) and (6) that 
\begin{eqnarray}
{d{\bf \rho }\over dt}&=&{c\over |{\bf B}|^2}
\left\{{\bf B\times }\big({\bf E}
+({\bf v\times B})/c \big)\right\}\nonumber \\
&=&\left({c{\bf B\times E}\over |{\bf B}|^2}\right)+{\bf v}_\perp 
\end{eqnarray}
and in virtue of Eqs.(6) and (7) we have the desired result 
\begin{equation}
{d{\bf R}\over dt}=\left({c{\bf E\times B}\over |{\bf B}|^2}\right).
\end{equation}

As discussed in a more leisurely fashion in previous work\cite{9}, 
the curvature length  
\begin{math} |{\rho }|=(cM\gamma |{\bf v}_\perp |/|e{\bf B}|) \end{math} 
of the muon path is the usual one for experimentally determining 
the muon transverse momentum. The coordinate 
\begin{math} {\bf R} \end{math} locates the center of the cyclotron 
orbit. The center \begin{math} {\bf R} \end{math} of the cyclotron 
orbit can also drift if an electric field acts on the muon charge in 
accordance with Eq.(8).

In fact, the entire horizontal electric field correction in Eq.(3)  
is due to the motion of the cyclotron orbit center velocity in Eq.(8). 
The classically predicted frequency shift is {\em exactly} given by  
\begin{equation}
{\delta \omega_{\perp \kappa} \over \omega_\kappa }={1\over c^2}
\left<\left\{{1\over \kappa (\gamma^2-1)}-1\right\}
{\bf v}_\perp \cdot \left({d{\bf R}\over dt}\right)\right>.
\end{equation}
The velocity \begin{math} {\bf v}_\perp  \end{math} circulates 
with a cyclotron angular velocity 
\begin{math} \omega_c =|e{\bf B}/Mc\gamma | \end{math}, 
while the center of the orbit \begin{math} {\bf R} \end{math}
circulates with an angular velocity 
\begin{math} \Omega_\perp \end{math} of the horizontal tune; 
Experimentally\cite{8}, 
\begin{equation}
(\Omega_\perp /\omega_c)\approx 7.0\times 10^{-2}. 
\end{equation}
In any case, the classical time average in Eq.(9) involves 
the two frequencies 
\begin{math} 
\left<\cos\{(\omega_c \pm \Omega_\perp )t+\Theta\}\right>\approx 0 
\end{math} 
leading to a vanishing frequency shift 
\begin{math} 
(\delta \omega_{\perp \kappa }/\omega_\kappa )\approx 0 
\end{math}.

That the electric field components \begin{math} {\bf E}_\perp \end{math} 
perpendicular to \begin{math} {\bf B}  \end{math} do not induce an 
anomaly frequency shift can be made abundantly clear by employing  
quantum theory via the Dirac equation for the muon. For electrostatic 
\begin{math} {\bf E}=-{\bf grad}\Phi  \end{math} and magnetostatic 
\begin{math} {\bf B}=curl{\bf A}  \end{math} fields and with 
the gauge invariant momentum 
\begin{math} {\bf \Pi}=-i\hbar {\bf grad}-(e/c){\bf A} \end{math}, 
the Dirac Hamiltonian is given by 
\begin{equation}
{\cal H}=c{\bf \alpha \cdot \Pi }+\beta Mc^2 +e\Phi +H_\kappa ,
\end{equation} 
where 
\begin{equation}
H_\kappa =
-\left({\hbar e\kappa \over 2Mc}\right)(\beta {\bf \Sigma \cdot B}
-i\beta \gamma_5{\bf \Sigma \cdot E}).
\end{equation} 

The discussion of the Dirac equation will proceed in 
three steps: (i) We consider the case in which only the magnetic field 
is present\cite{10}. For motion in a plane perpendicular to the magnetic 
field, the anomaly frequency in Eq.(2) arises because the operator 
\begin{equation}
\left({\hbar \omega_\kappa \over 2}\right) \Lambda =
\left({\hbar e\kappa \over 2Mc}\right)\beta {\bf \Sigma \cdot B}=
\left({\hbar \omega_\kappa \over 2}\right)\beta \Sigma_3.  
\end{equation}
is conserved. (ii) We now add the electric field components 
\begin{math} {\bf E}_\perp \end{math} and prove 
that \begin{math} \Lambda \end{math} is again conserved for planar 
motions. Thus, the frequency shift 
\begin{math} \delta \omega_{\perp \kappa } \end{math} due to 
\begin{math} {\bf E}_\perp \end{math} is then {\em strictly zero}. 
This is a {\em rigorous} quantum selection rule. (iii) When the field 
component 
\begin{math} {\bf E}_{||} \end{math}
parallel to the magnetic field is included there will be a frequency 
shift due to normal (out of the plane) oscillations in the muon beam.
The implications of these results for the experimental determination 
of \begin{math} \kappa \end{math} will be discussed below.

In zero electric field and a uniform magnetic field with 
\begin{math} {\bf A}=(1/2){\bf B\times r} \end{math}, 
the Dirac Hamiltonian reads 
\begin{eqnarray}
{\cal H}_{\bf B}(p_z)&=&c\alpha_3 p_z
+c{\bf \alpha_\perp \cdot \Pi_\perp }
+\beta Mc^2 -(\hbar \omega_\kappa /2)\Lambda \nonumber \\
&=&H_{\bf B}(p_z)-(\hbar \omega_\kappa /2)\Lambda .
\end{eqnarray}
For motion in the plane, the operator 
\begin{math} \Lambda =\beta \Sigma_3  \end{math}
is conserved 
\begin{equation}
\left[{\cal H}_{\bf B}(p_z=0),\Lambda \right]=0.
\end{equation}
One may diagonalize the ``in plane'' motion according to 
\begin{math} H_{\bf B}(p_z=0)\psi_{{\cal E}\pm}={\cal E}
\psi_{{\cal E}\pm } \end{math} 
and \begin{math} 
\Lambda \psi_{{\cal E}\pm}=\pm \psi_{{\cal E}\pm } 
\end{math}.
Thus, the in plane motion energy states obey 
\begin{equation}
{\cal H}_{\bf B}(p_z=0)\psi_{{\cal E}\pm}=
\{{\cal E}\mp (\hbar \omega _\kappa /2)\}\psi_{{\cal E}\pm}.
\end{equation}
\medskip
\par \noindent 
{\bf Theorem I:} {\em For planar motions in a uniform magnetic field, 
the Bohr transition frequency 
\begin{math} \omega_\kappa =({\cal E}_--{\cal E}_+)/\hbar  \end{math}
for chiral spin rotations is precisely the anomaly frequency of} Eq.(2).
\medskip 

Let us now consider the theoretical case where, in addition to the 
uniform magnetic field \begin{math} {\bf B} \end{math}, there exists 
an in plane electric field 
\begin{math}
{\bf E}_\perp =-\nabla_\perp \Phi_\perp ({\bf r}_\perp ).
\end{math} 
For this case Eq.(14) is modified to read 
\begin{eqnarray}
{\cal H}_{{\bf B,E}_\perp }(p_z)&=&H_{{\bf B,E}_\perp }(p_z)
-(\hbar \omega_\kappa /2)\Lambda \nonumber \\
H_{{\bf B,E}_\perp }(p_z)&=&c\alpha_3 p_z
+c{\bf \alpha_\perp \cdot \Pi_\perp }+\beta Mc^2 \nonumber \\  
&+&e\Phi_\perp +\left({\hbar e\kappa \over 2Mc}\right)
i\beta \gamma_5{\bf \Sigma}_\perp {\bf \cdot E}_\perp .
\end{eqnarray}
Remarkably, for planar motions the operator 
\begin{math} \Lambda =\beta \Sigma_3  \end{math}
is still conserved 
\begin{equation}
\left[{\cal H}_{{\bf B,E}_\perp }(p_z=0),\Lambda \right]=0.
\end{equation}
Again, one may diagonalize the in plane motion according to 
\begin{math} 
H_{{\bf B,E}_\perp }(p_z=0)\psi _{\tilde{\cal E} \pm}=
\tilde{\cal E}
\psi _{\tilde{\cal E} \pm} 
\end{math} 
and 
\begin{math} 
\Lambda \psi _{\tilde{\cal E} \pm}=\pm \psi _{\tilde{\cal E} \pm} 
\end{math} so that 
\begin{equation}
{\cal H}_{{\bf B,E}_\perp }(p_z=0)
\psi _{\tilde{\cal E} \pm}=
\{\tilde{\cal E} \mp (\hbar \omega _\kappa /2)\}
\psi _{\tilde{\cal E} \pm}.
\end{equation}
\medskip
\par \noindent 
{\bf Theorem II:} {\em For planar motions with both a uniform magnetic 
field and any in plane electric field, 
the Bohr transition frequency 
\begin{math} 
\omega_\kappa =(\tilde{\cal E}_--\tilde{\cal E}_+)/\hbar  
\end{math}
for chiral spin rotations is precisely the anomaly frequency of} Eq.(2) 
{\em with a strictly zero frequency shift}.

The above theoretical quantum mechanical proof of 
\begin{math} \delta \omega_{\perp \kappa} =0 \end{math}, 
employing the conservation law for the operator 
\begin{math} \Lambda=\beta \Sigma_3  \end{math}, 
is a central result of this work. 
For the full Hamiltonian in Eqs.(11) and (12), 
\begin{math} \Lambda \end{math} is not quite 
conserved. In detail,
\begin{math}
\dot{\Lambda }=(i/\hbar)\left[{\cal H},\Lambda \right]
\end{math} 
is given by 
\begin{equation}
\dot{\Lambda }=-\left({2cp_z\over \hbar }\right)i\beta \gamma_5 
+\left({\kappa eE_z\over Mc}\right)\gamma_5.
\end{equation}
The {\em only frequency shift that can arise} is due solely to 
non-planar motions with momentum 
\begin{math}
p_z=({\bf p}\cdot{\bf B})/|{\bf B}|
\end{math}
out of the plane driven by the electric field component  
\begin{math}
E_z=({\bf E}\cdot{\bf B})/|{\bf B}|
\end{math}. 
The vertical betatron oscillation correction is thus a bit more subtle.
%%%%%%%%%%%%%%%%%%%%%%%%%%%%%%%%%%%%%%%%%%%%%%%%%%%%%%%%%%%%%%%%%%%%%%%%

%%%%%%%%%%%%%%%%%%%%%%%%%%%%%%%%%%%%%%%%%%%%%%%%%%%%%%%%%%%%%%%%%%%%%%%% 
From a classical relativistic viewpoint, the planar motions of the muons 
may be considered to be within a circulating ``hoop'' with an ``effective 
mass'' 
\begin{equation} 
M_{eff}c^2=\tilde{\cal E}. 
\end{equation}
The effective classical Lagrangian for the vertical betatron 
oscillation in the \begin{math} z \end{math}-direction may then 
be shown to be 
\begin{equation}
L_{||}(v_z,z)=-\tilde{\cal E}\sqrt{1-\left({v_z\over c}\right)^2}
-e\Phi_{||}(z),
\end{equation}
where the electric field \begin{math} E_z=-(d\Phi_{||}(z)/dz)  \end{math} 
describes the restoring force of the vertical betatron oscillation at 
frequency
\begin{equation}
\Omega_{||}=\sqrt{n}\omega_c
\end{equation}
where \begin{math} n \end{math} is the electric field index. The 
betatron oscillator Lagrangian in Eq.(22) renormalizes the 
anomaly frequency according to the relativistic Doppler frequency shift 
formula 
\begin{equation}
\hbar \bar{\omega }_\kappa =\left<
(\tilde{\cal E}_--\tilde{\cal E}_+)
\sqrt{1-\left({v_z\over c}\right)^2}\ \right>,
\end{equation}
i.e. the classical (transverse) Doppler shift for 
\begin{math} |v_z|<<c \end{math} 
is given by 
\begin{equation}
{\delta \omega_{||\kappa }\over \omega_\kappa}\approx 
-{1\over 2}\left<\left({v_z\over c}\right)^2\right>\approx 
-\left({\Omega_{||}^2\over 2c^2}\right)\left<z^2\right>. 
\end{equation}

An alternative quasi-classical argument for the vertical 
betatron induced shift reads as follows: (i) The number 
\begin{math} N>>1  \end{math} of ``quanta'' in the vertical betatron 
oscillation for a total energy \begin{math} {\cal E} \end{math} and 
transverse energy \begin{math} \tilde{\cal E} \end{math} 
is given by the Bohr rule 
\begin{eqnarray}
N&=&\oint \left({p_zdz\over 2\pi \hbar}\right), \nonumber \\
cp_z&=&\sqrt{({\cal E}-e\Phi_{||}(z))^2-\tilde{\cal E}^2}.
\end{eqnarray}
(ii) The velocity  
\begin{math}
v_z=(\partial {\cal E}/\partial p_z)_{\tilde{\cal E},z}
\end{math}
determines the vertical betatron tune frequency via 
\begin{eqnarray}
\left(\partial N\over \partial {\cal E}\right)_{\tilde{\cal E}}
&=&\oint \left(
{\partial p_z\over \partial \cal E}
\right)_{\tilde{\cal E},z}\left({dz\over 2\pi \hbar }\right), 
\nonumber \\
\left(\partial N\over \partial {\cal E}\right)_{\tilde{\cal E}}
&=&\oint \left({dz\over 2\pi \hbar v_z}\right)=
\left({1 \over \hbar \Omega_{||}}\right), \nonumber \\
\left(\partial N\over \partial \tilde{\cal E}\right)_{\cal E}
&=&-\oint {\tilde{\cal E}\over {\cal E}-e\Phi_{||}(z)} 
\left({dz\over 2\pi \hbar v_z}\right).
\end{eqnarray}
The maximum vertical betatron amplitude 
\begin{math} z_0  \end{math} obeys 
\begin{equation}
{\cal E}=\tilde{\cal E}+ e\Phi_{||}(z_0)-e\Phi_{||}(0),
\end{equation}
and without loss of generality we may choose 
\begin{math} \Phi_{||}(0)=0 \end{math}.
Eqs.(27) and (28) then imply 
\begin{eqnarray}
\left({\partial {\cal E}\over \partial \tilde{\cal E}}\right)_N
&=& -\left({\partial N\over \partial \tilde{\cal E}}\right)_{\cal E}
\left({\partial {\cal E}\over \partial N}\right)_{\tilde{\cal E}}
\nonumber \\
\left({\partial {\cal E}\over \partial \tilde{\cal E}}\right)_N &=& 
\left<{\tilde{\cal E}
\over \tilde{\cal E}+e\{\Phi_{||}(z_0)-\Phi_{||}(z)\}}\right>,
\end{eqnarray}
where the average is over the vertical betatron oscillation.
The frequency shift  
\begin{equation}
\left({\delta \omega_{||\kappa }\over \omega_\kappa }\right)=1-
\left({\partial {\cal E}\over \partial \tilde{\cal E}}\right)_N
\end{equation}
thereby obeys 
\begin{equation}
\left({\delta \omega_{||\kappa }\over \omega_\kappa }\right)=
\left<{e\{\Phi_{||}(z)-\Phi_{||}(z_0)\}
\over \tilde{\cal E}+e\{\Phi_{||}(z_0)-\Phi_{||}(z)\}}\right>.
\end{equation}
For an oscillator with 
\begin{math} \tilde{\cal E}>>|e\{\Phi_{||}(z_0)| \end{math}
Eq.(31) reads 
\begin{equation}
{\delta \omega_{||\kappa }\over \omega_\kappa}\approx  
-\left({\Omega_{||}^2\over 2c^2}\right)\left<z^2\right>, 
\end{equation}
in agreement with Eq.(25). Finally, Eq.(32) yields the standard 
vertical betatron frequency shift formula 
\begin{equation}
\left({\delta \omega_{||\kappa }\over \omega_\kappa}\right)
\approx -\left({n\over 2\rho_0^2}\right)\left<z^2\right>
\approx -0.23\times 10^{-6}.
\end{equation}
since \begin{math} n\approx 0.137 \end{math} and the cyclotron 
radius \begin{math} \rho_0\approx 711.2\ cm \end{math}
\cite{7,8}.

\begin{figure}[bp]
\scalebox {0.5}{\includegraphics{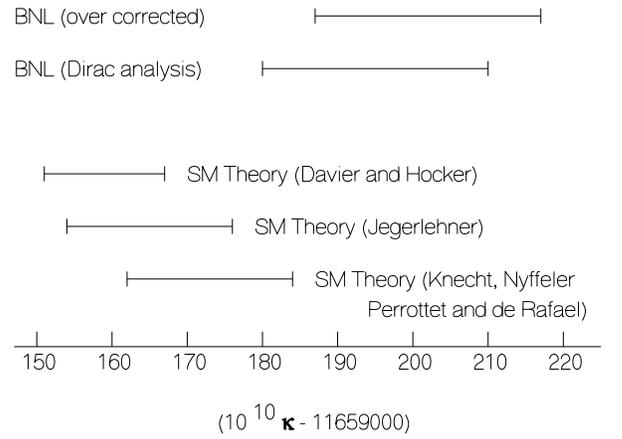}}
\caption{The value of $\kappa =(g-2)/2$ is shown together with estimated 
errors. Reading from top to bottom: (i) the experimental BNL data with 
the ``over compensation'' of the electric field corrections as reported 
in \cite{7}, (ii) the experimental data as analyzed employing the quantum 
Dirac equation of this work and thereby including only the vertical 
betatron tune, (iii) the theoretical standard model calculation as 
reported in \cite{11}, (iv) the theoretical standard model calculation 
as reported in \cite{12} and (v) the theoretical standard model calculation 
as reported in \cite{13,14}.}
\label{fig1}
\end{figure}

Shown in Fig.1 are the central values and error ranges for the 
BNL experimental measurement of \begin{math} \kappa  \end{math} 
as well as the more recent {\em standard model theory} values 
and error ranges. 

The original experimental report over-compensated 
the electric field correction to \begin{math} \omega_\kappa \end{math}. 
The electric field \begin{math} {\bf E}_\perp \end{math} contributes 
zero to a frequency shift as shown above employing the Dirac 
wave function for the muon beam. From a classical viewpoint, when the 
muons are first injected into the beam line, the ``kick'' slightly 
off-centers the orbit setting the center 
\begin{math} {\bf R} \end{math} 
of the cyclotron orbit to rotate at the horizontal betatron 
frequency \begin{math} \Omega_\perp \end{math}. This effect on 
\begin{math} \omega_\kappa  \end{math} averages 
away to zero. The Dirac wave function analysis thus yields a lower 
magnitude of the electric field shift than was previously reported.  
The resulting \begin{math} \kappa \end{math} with the Dirac 
equation analysis is pictured in Fig.1. This describes   
the only possible electric field correction to  
\begin{math} \omega_\kappa  \end{math} as given by Eq.(33)

The damping of the horizontal betatron mode velocity 
\begin{math} {\bf V}=(d{\bf R}/dt) \end{math} takes place in an 
estimated  time of \begin{math}\sim 168\ microsec \end{math}.
In this regard, if the damping of the vertical betatron coordinate 
\begin{math} z  \end{math} takes place in a comparable time, then 
the vertical betatron contribution Eq.(33) may also be moot. 

The nature and values of the direct vertical betatron 
measurements have not yet been made available to the physics 
community. If the vertical tune relaxation times are sufficiently small,  
then the standard model theory and the  
\begin{math} (g-2) \end{math} experiment would be brought 
into {\em yet closer agreement} than is shown in Fig.1.

\end{document}